\documentclass[amssymb,aps,twocolumn,prb]{revtex4-2}
\usepackage{epsfig,subfigure,amsmath,float,xcolor,ulem}
\usepackage[colorlinks]{hyperref}
\hypersetup{colorlinks=true,linkcolor=blue,filecolor=blue,urlcolor=blue,citecolor=blue}
\usepackage[utf8]{inputenc}
\usepackage{soul}
\usepackage{comment}
 \usepackage{graphicx}
\usepackage{physics}

\newcommand\beq{\begin{equation}}
\newcommand\eeq{\end{equation}}
\newcommand\bes{\begin{subequations}}
\newcommand\ees{\end{subequations}}
\newcommand\bea{\begin{eqnarray}}
\newcommand\eea{\end{eqnarray}}

\newcommand\ig{\includegraphics}

\newcommand\al{\alpha}
\usepackage{yfonts}

\newcommand\be{\beta}
\newcommand\ga{\gamma}

\newcommand\lam{\lambda}

\newcommand\om{\omega}

\newcommand\sfig{\subfigure}

\newcommand\ver{{\bf r}}

\newcommand\vk{{\bf k}}

\begin{document}

\title{Chemical potential of magnon polarons}
\author{Violet Williams}
\email{willibte@bc.edu}
\affiliation{Department of Physics, Boston College, 140 Commonwealth Avenue Chestnut Hill,
Massachusetts 02467, USA}
\author{Benedetta Flebus}
\email{flebus@bc.edu}
\affiliation{Department of Physics, Boston College, 140 Commonwealth Avenue Chestnut Hill,
Massachusetts 02467, USA}
\date{\today}
\begin{abstract}
    While magnons have long been regarded as the sole carriers of angular momentum in magnetic insulators, a growing body of theory and experiment has revealed that acoustic phonons can also transport angular momentum—intrinsically in certain crystals and, more generally, through magnon–phonon hybridization. This realization calls for a framework in which spin and lattice angular momentum are treated on equal footing, requiring a thermodynamically consistent redefinition of the chemical potential tied to the quasi-conserved axial angular momentum once magnons and phonons hybridize into magnon–polaron modes. Using a rotationally invariant formulation of spin–lattice coupling, here we derive a rigorous definition of the chemical potential for magnon–polaron quasiparticles in collinear ferromagnets (FMs) and antiferromagnets (AFMs)—valid when magnetoelastic scattering equilibrates magnons and acoustic phonons on timescales much shorter than those associated with quasiparticle–nonconserving relaxation processes. While our approach is general, here we focus on high–symmetry crystals where transverse acoustic modes combine into circularly polarized phonons carrying opposite angular-momentum components. In this basis, the chiral selectivity of magnetoelastic coupling becomes explicit: the FM magnon mode hybridizes only with the co-rotating phonon, whereas in a collinear AFM each magnon branch of opposite handedness couples to the phonon of matching chirality.
    We show that, in both cases, the resulting nonequilibrium magnon–polaron gas is governed by a single chemical potential conjugate to the conserved axial angular momentum. In FMs, the two hybrid branches within the co-rotating sector share this potential according to their magnonic weight; in AFMs, the four hybrid branches organize into two chiral sectors that carry opposite angular momenta and couple with opposite sign to the same potential. Building on this structure, we develop a Boltzmann transport theory for magnon–polarons and derive compact expressions for angular-momentum and heat currents that recover the expected limits as magnetoelastic coupling is tuned to zero. Our work places the notion of a magnon–polaron chemical potential on firm microscopic footing and provide a systematic route for incorporating angular-momentum conservation into transport theories of magnetic insulators.
\end{abstract}

\maketitle

\section{Introduction}
In the past decade, the concept of a magnon chemical potential has become central to the description of spin and energy transport in magnetically ordered insulators~\cite{Flebus2021Magnonics}. It is now well established that, when spin–nonconserving interactions are much weaker than the exchange energy, the total magnon number can be treated as approximately conserved~\cite{bennett2008chemical}. In this regime—relevant to both thermally driven spin currents and magnon Bose–Einstein condensation~\cite{demokritov2006bose}—the magnon chemical potential emerges as a natural, and in practice essential, thermodynamic parameter  characterizing deviations from equilibrium~\cite{Cornelissen2016Magnon,Demidov2017Chemical,du17,hoogeboom2020nonlocal,olsson2020pure}. 

Implicit in this description is the assumption that magnons alone carry spin angular momentum, with phonons playing solely the role of a passive thermal reservoir. 
Recent work has, however, challenged this view: in a broad class of systems—such as noncentrosymmetric or ionic crystals—acoustic phonons carry an intrinsic lattice angular momentum, and in magnetic materials they can acquire angular momentum via spin–orbit–mediated coupling to the ordered spins~\cite{zhang2014angular,coh2023classification,flebus2023phonon,juraschek2025chiral}.

These discoveries have prompted a reassessment of standard phenomenological models and led to  the development of rotationally invariant formulations of spin–lattice coupling that explicitly enforce global angular momentum conservation~\cite{mankovsky2022angular,weissenhofer2023rotationally,ruckriegel2020angular}. In contrast to conventional symmetry–based magnetoelastic theories~\cite{birss1966symmetry,callen1963static,brown1965theory}, these approaches provide a microscopically consistent description of angular–momentum exchange between spin and lattice degrees of freedom,
 which has yielded key insights into phenomena ranging from ultrafast demagnetization dynamics~\cite{tauchert2022polarized,davies2024phononic,Dornes2019,hennecke2019angular,mrudul2025generation} to the chiral selectivity of magnon–phonon interactions~\cite{cui2023chirality,weiss24}.

While the implications of angular–momentum conservation for coherent spin–lattice dynamics are becoming increasingly clear, its consequences for incoherent spin transport remain largely unexplored. This question becomes particularly relevant when the rate at which magnons and phonons exchange energy and thermalize is much faster than their intrinsic decay rates,  so that magnons and phonons cease to act as independent carriers and instead form hybrid magnon–polaron quasiparticles~\cite{kittel1958interaction,Gode20,sim19,Lij20,streib21,sukh19}.

A growing body of experiments indicates that magnon–phonon hybridization plays a central role in thermally driven spin transport. For instance, field–dependent anomalies in spin Seebeck voltages---appearing as characteristic peaks or dips---have been consistently linked to hybridization between magnons and phonons with distinct scattering rates~\cite{flebus17,kikkawa2016magnon,zhang2021long,wang18,Li24}. This interpretation is reinforced by the observation that these anomalies are most pronounced at magnetic fields where hybridization is maximized, and that their character can be tuned by adjusting the relative magnetic and acoustic quality of the material~\cite{kikkawa22,Ramos19,Yang21}.
This phenomenological picture implicitly assumes that phonons, once hybridized with magnons, actively contribute to spin transport and relies  on the notion of an effective magnon–polaron chemical potential. To date, however, this quantity has been introduced heuristically, i.e., within a theoretical framework that does not explicitly enforce angular–momentum conservation~\cite{flebus17}. While these models successfully reproduce key experimental trends, they leave open the fundamental question of whether---and how---this thermodynamic variable can be rigorously defined for hybrid magnon–phonon excitations.

Here we address this conceptual gap by establishing---within a microscopic, rotationally invariant description of spin–lattice angular–momentum exchange---a definition of the magnon–polaron chemical potential in collinear ferromagnetic and antiferromagnetic insulators, tied to a well–defined, quasi-conserved spin-like component of the total angular momentum.
We focus on a high--symmetry crystal in which the two transverse acoustic (TA) phonon modes remain degenerate and combine into circularly polarized states carrying angular momentum \(\pm\hbar\) along the equilibrium magnetization. While our framework relies only on the quasi-conservation of the total axial angular momentum of the coupled spin--lattice system and it is thus basis independent, this high–symmetry example makes the chiral selectivity of the interaction particularly transparent.

In collinear ferromagnets, the single magnon mode hybridizes exclusively with the co-rotating phonon branch. We find that, as a result, the two resulting magnon–polaron branches share a common chemical potential, weighted by their respective magnonic spectral content.  In collinear antiferromagnets, the two magnon helicities hybridize with the correspondingly chiral phonons, resulting in four magnon–polaron bands that organize into two decoupled chiral sectors. We show that the four modes are governed by a single chemical potential associated with the conserved angular momentum; this parameter couples with opposite sign to the two sectors and with strength set by their magnonic content. 
In the limit of vanishing hybridization, our results continuously reduce to the known expressions for the FM and AFM magnon chemical potential~\cite{flebus19,Cornelissen2016Magnon}.

Building on these insights, we develop a Boltzmann transport theory for FM and AFM magnon–polaron quasiparticles and derive explicit expressions for the associated spin and heat currents in terms of the magnonic spectral weights and scattering rates of each hybrid branch. In the ferromagnetic case, we recover exactly the magnon–polaron spin and heat current expressions reported in Ref.~\cite{flebus17}.

This work is organized as follows. In Sec.~\ref{sec:model}, we introduce the magnon and phonon Hamiltonians and derive the magnon–phonon coupling within a rotationally invariant framework. In Sec.~\ref{sec:magpol1}, we define the hybrid magnon–polaron quasiparticles and their total angular momentum. In Sec.~\ref{sec:transport}, we obtain their chemical potentials and develop a linearized Boltzmann theory for angular–momentum and heat transport. In Sec.~\ref{sec:con}, we summarize our main results and outline prospects for future work.

\section{Model} \label{sec:model}
In this Section, we introduce the spin and lattice Hamiltonians for collinear ferromagnetic and antiferromagnetic insulators and derive the corresponding rotationally invariant spin–lattice coupling. Throughout, we work with a coarse–grained low–energy description in which the magnetic insulator is represented by an effective cubic lattice, and the elastic medium is treated as isotropic with one longitudinal and a degenerate transverse acoustic branch. This approximation does not confine our analysis to strictly cubic crystals, as it is routinely used for a wide range of FM and AFM systems, whose long–wavelength dynamics are well captured by an effective cubic, nearly isotropic continuum. 

\subsection{Phonons} \label{sec:phonon}

We consider an isotropic elastic solid, whose acoustic phonon Hamiltonian is expressed in terms of phonon creation and annihilation operators, \(c^{\dagger}_{\vk,\lambda}\) and \(c_{\vk,\lambda}\), as
\beq
\mathcal{H}_{\text{p}} = \sum_{\vk,\lam}\hbar \om_{\text{p},\lam}(\vk) \Bigg(c^\dag_{\vk, \lam} c_{\vk,\lam}+\frac{1}{2}\Bigg),
\eeq
where $\lambda\in\{1,2,\parallel \}$ labels the two transverse acoustic (TA) branches and the longitudinal acoustic (LA) branch, respectively. For an isotropic medium, the frequencies are linear, i.e., 
$
\omega_{\mathrm{p},\lambda}(\mathbf{k}) = v_\lambda \lvert \mathbf{k} \rvert
$
with sound velocities \(v_\lambda\) depending on polarization.
The corresponding unit polarization vectors can be chosen as
\begin{align}
\boldsymbol{\epsilon}_{1}(\mathbf{k}) &= \bigl(\cos\theta_{\mathbf{k}}\cos\phi_{\mathbf{k}},\ \cos\theta_{\mathbf{k}}\sin\phi_{\mathbf{k}},\ -\sin\theta_{\mathbf{k}}\bigr), \nonumber\\
\boldsymbol{\epsilon}_{2}(\mathbf{k}) &= \bigl(-\sin\phi_{\mathbf{k}},\ \cos\phi_{\mathbf{k}},\ 0\bigr), \nonumber\\
\boldsymbol{\epsilon}_{\parallel}(\mathbf{k}) &= \bigl(\sin\theta_{\mathbf{k}}\cos\phi_{\mathbf{k}},\ \sin\theta_{\mathbf{k}}\sin\phi_{\mathbf{k}},\ \cos\theta_{\mathbf{k}}\bigr),
\label{eq:polarizations}
\end{align}
which form an orthonormal triad that satisfies
\begin{align}
\boldsymbol{\epsilon}_{\lambda}(\mathbf{k})\cdot \boldsymbol{\epsilon}_{\lambda'}(\mathbf{k})=\delta_{\lambda\lambda'}
\quad\text{and}\quad
\boldsymbol{\epsilon}_{1}(\mathbf{k})\times \boldsymbol{\epsilon}_{2}(\mathbf{k})=\hat{\mathbf{k}}\,.
\end{align} 
The quantized lattice displacement field reads as
\begin{equation}
    \mathbf{u}(\mathbf{r})  = \sum_{\mathbf{k},\lambda}\sqrt{\frac{\hbar}{2\Bar{\rho}V}} \frac{e^{-i\mathbf{k}\cdot \mathbf{r}}}{\sqrt{\omega_{\text{p},\lambda}(\mathbf{k})}}\boldsymbol{\epsilon}_\lambda(\mathbf{k}) (c^\dag_{\vk, \lam} + c_{-\vk,\lam})\,,
    \label{eq:uquant}
\end{equation}
where \(V\) is the crystal volume and \(\bar{\rho}\) the average mass density.
A central quantity to this work is the phonon angular momentum $\mathbf{L} = \int d^3r \bar{\rho}\, \mathbf{u}\times \dot{\mathbf{u}}$~\cite{zhang2024observation}, which, using Eq.~\eqref{eq:uquant}, can be rewritten as
\begin{equation}
\mathbf{L} = -\frac{i\hbar}{2}
\sum_{\mathbf{k},\lambda \ne \lambda'}
\bigl(\boldsymbol{\epsilon}_{\lambda}(\mathbf{k})\times \boldsymbol{\epsilon}_{\lambda'}(\mathbf{k})\bigr)
\left(
c^\dagger_{\mathbf{k},\lambda}c_{\mathbf{k},\lambda'}
-
c^\dagger_{\mathbf{k},\lambda'}c_{\mathbf{k},\lambda}
\right).
\label{eq:L2linear}
\end{equation}

Linearly polarized phonons (e.g., LA modes) do not carry angular momentum, as their lattice displacements lack a definite sense of rotation~\cite{ren22}. By contrast, circularly or elliptically polarized phonons can host intrinsic angular momentum, encoded in the rotation of the atomic displacement vector about the propagation direction~\cite{juraschek2025chiral}. In high-symmetry crystals with twofold-degenerate transverse acoustic branches throughout the Brillouin zone -- for instance, cubic systems with \(v_1 = v_2\) -- the TA doublet spans, for each \(\mathbf{k}\), an eigenspace of the elastic Hamiltonian. Thus, the corresponding circularly polarized combinations 
\begin{equation}
c_{\mathbf{k},\pm}
= \frac{1}{\sqrt{2}}\bigl(c_{\mathbf{k},1} \pm i c_{\mathbf{k},2}\bigr),
\label{eq:defcirc}
\end{equation}
are themselves exact eigenmodes with polarization
\(
\boldsymbol{\epsilon}_{\pm}(\mathbf{k}) =
\boldsymbol{\epsilon}_{1}(\mathbf{k}) \mp i \boldsymbol{\epsilon}_{2}(\mathbf{k}),
\)
carrying angular momentum \(\mp \hbar\).
Since the polarization vectors form an orthonormal triad, such that
$\boldsymbol{\epsilon}_{\lambda}(\mathbf{k}) \times 
\boldsymbol{\epsilon}_{\lambda'}(\mathbf{k}) = \pm \mathbf{k}/k$,
Eq.~\eqref{eq:L2linear} can be rewritten as
\begin{equation}
\mathbf{L} = -\hbar \sum_{\mathbf{k}}\frac{\mathbf{k}}{k}
\bigl(c^\dagger_{\mathbf{k},+}c_{\mathbf{k},+}
- c^\dagger_{\mathbf{k},-}c_{\mathbf{k},-}\bigr)\,.
\label{eq:L2}
\end{equation}

Equation~(\ref{eq:L2}) shows that the phonon angular momentum is set by the population imbalance between left– and right–handed circularly polarized modes. In time–reversal–symmetric  crystals these chiral branches are exactly degenerate, so their equilibrium populations are equal and the net angular momentum vanishes. Breaking time–reversal symmetry  (TRS) lifts this degeneracy and allows phonons to carry a finite angular momentum. In the following, we show how coupling to a magnetic subsystem provides a natural source of TRS breaking and redistributes angular momentum between the lattice and the spin degrees of freedom.

\begin{figure}[htb]
\sfig[]{\ig[width = 8cm,trim={2cm 2cm 2cm 2cm},clip]{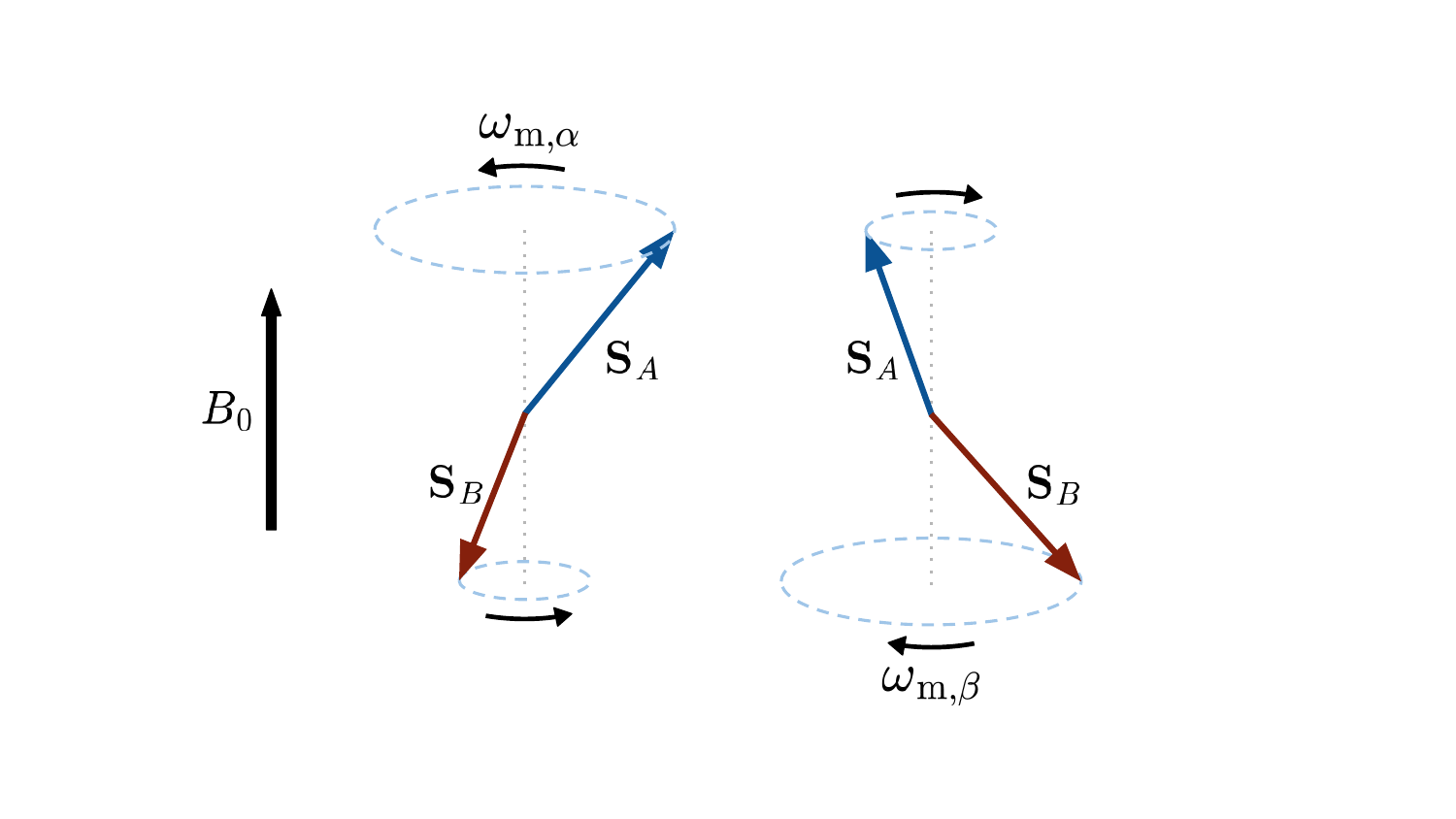}\label{fig:schematic1}}
\sfig[]{\ig[width = 8.8cm,trim={3.5cm 4cm 4.3cm 3cm},clip]{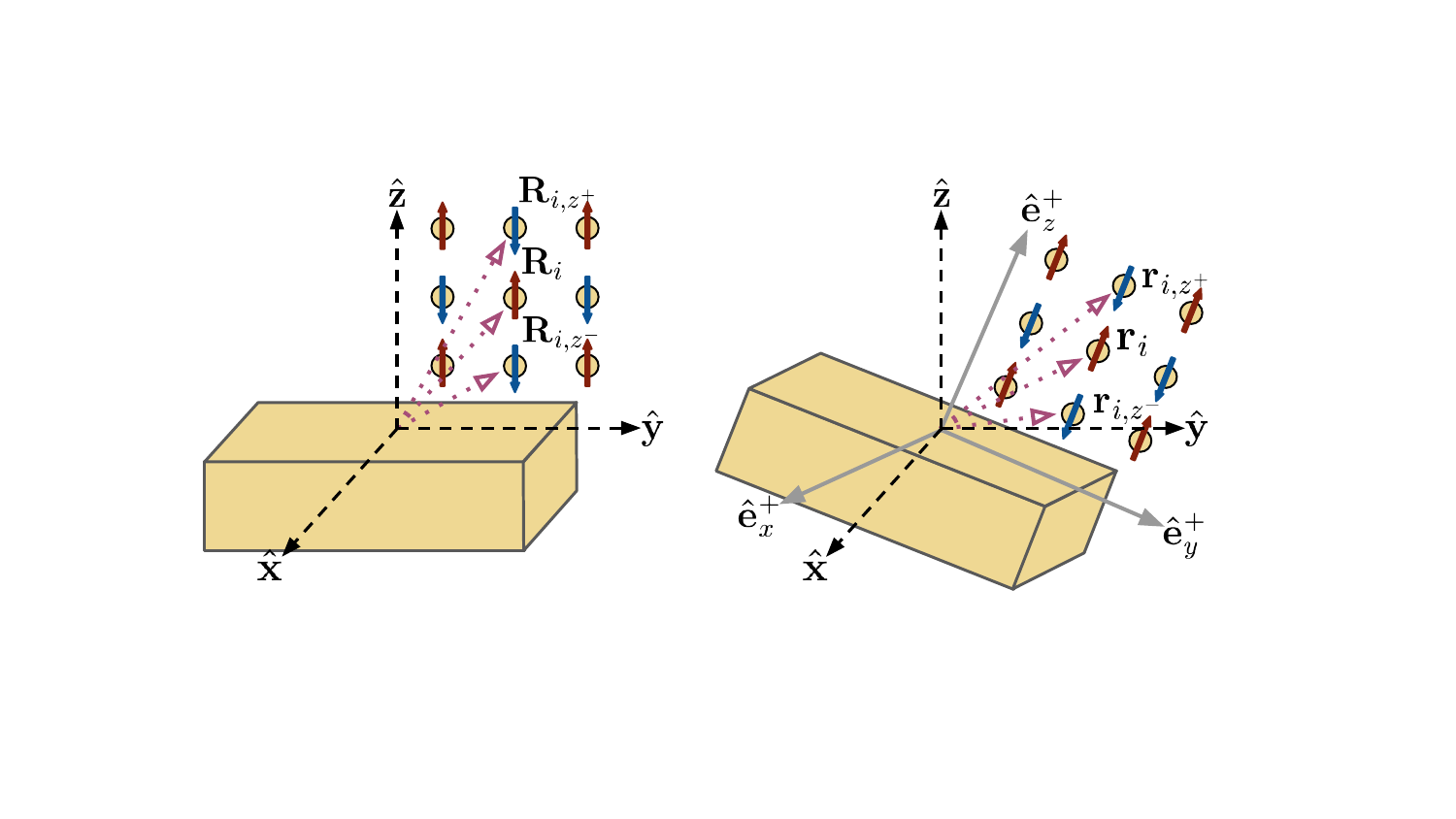}\label{fig:schematic2}}
\caption{Schematic illustration of the magnon eigenmode precession and the rotationally invariant construction used to define the magnetoelastic coupling in a collinear antiferromagnet.
(a) Linear spin–wave eigenmodes of the two–sublattice antiferromagnet. The \(\alpha\) and \(\beta\) modes correspond to counter-rotating collective precessions of the sublattices and carry angular momentum \(-\hbar\) and \(+\hbar\), respectively, forming the natural chiral basis in which the antiferromagnetic BdG Hamiltonian is diagonal.
(b) Equilibrium and displaced positions \(\mathbf{R}_i\) and \(\mathbf{r}_i\) of a G-type antiferromagnet. The local easy–axis at site \(i\) is defined geometrically from the instantaneous positions of the nearest neighbors above and below the site, \(\mathbf{r}_{i,\,z\pm}\), thereby ensuring that the anisotropy axis \(\hat{\mathbf{e}}^{\pm}_{z}(\mathbf{r}_i)\) simultaneously rotates with lattice distortions and preserves global rotational invariance.}
\end{figure}

\subsection{Magnons} \label{sec:AFM}

Turning to the magnetic sector, we proceed in two steps: we first derive the magnon spectrum from a conventional microscopic spin Hamiltonian, and then restore rotational invariance to obtain the magnon–phonon coupling. We consider a magnetic insulator whose spins $\mathbf{S}_i \equiv \mathbf{S}(\mathbf{r}_i)$, with $ |\mathbf{S}_i|=S$, are localized at the lattice  sites $\mathbf{r}_i$. A minimal model for the spin dynamics is given by the Hamiltonian~\footnote{Further details on the derivation of the dipolar-interaction contributions are provided in the Supplemental Material.}:
\begin{equation}
\mathcal{H}_{\text{m}} = -\mathcal{J} \sum_{\langle ij \rangle} \mathbf{S}_i \cdot \mathbf{S}_j - K \sum_i (\mathbf{S}_i \cdot \hat{\textbf{z}})^2 - \gamma \hbar B_0 \sum_i S_i^z,
\label{eq:Hm}
\end{equation}
where \(\mathcal{J}\) is the nearest–neighbors Heisenberg exchange constant, $K>0$  the easy–axis anisotropy along the global $\hat{\textbf{z}}$ direction, $\gamma $  the gyromagnetic ratio, and $B_0$ the external magnetic field.
For \(\mathcal{J}>0\) the ground state is ferromagnetic, whereas for \(\mathcal{J}<0\) it is a collinear G-type antiferromagnet with two interpenetrating sublattices \(A\) and \(B\). 
Linear spin–wave theory is obtained by performing a Holstein–Primakoff expansion about the corresponding ordered state~\cite{hptransform}:
\begin{align}
S_{A,i}^+ &= \sqrt{2S}\, a_i\,, & S_{A,i}^z &= S - a_i^\dagger a_i\,, \label{AA}\\
S_{B,i}^+ &= \sqrt{2S}\, b_i^\dagger\,, & S_{B,i}^z &= -S + b_i^\dagger b_i\,,
\label{eq:hp1}
\end{align}
where \(a_i\) (\(a_i^\dagger\)) and \(b_i\) (\(b_i^\dagger\)) annihilate (create) magnons on sublattices \(A\) and \(B\), respectively, and obey standard bosonic commutation relations. In the FM case, the usual single–branch spin–wave expansion is recovered by applying the \(A\)-sublattice transformation~\eqref{AA} to all lattice sites \(i\).
Substituting the Fourier representation
\begin{equation}
a_i = \frac{1}{\sqrt{N}} \sum_\vk e^{i \vk \cdot \ver_i} a_{\vk}\,, \quad
b_i = \frac{1}{\sqrt{N}} \sum_\vk e^{i \vk \cdot \ver_i} b_{\vk}\,,
\end{equation}
with $N$ the number of lattice sites, and retaining only quadratic terms in the bosonic operators, the AFM Hamiltonian~\eqref{eq:Hm} reduces to
\begin{align}
&\mathcal{H}_{\mathrm{m}}^{\text{\tiny(AFM)}}
= \gamma\hbar \sum_{\mathbf{k}}
\Big[ D\,\gamma_{\mathbf{k}}\,\big(a_{\mathbf{k}}\, b_{-\mathbf{k}} + a_{\mathbf{k}}^\dagger b_{-\mathbf{k}}^\dagger\big)
 \nonumber \\
&+(D + A + B_0)\, a_{\mathbf{k}}^\dagger a_{\mathbf{k}}+ (D + A - B_0)\, b_{\mathbf{k}}^\dagger b_{\mathbf{k}}
\Big]\,,
\label{eq:HAFM}
\end{align}
where $D=2S z_0 |J|/(\gamma\hbar)$ is the exchange field, $A=2S |K|/(\gamma\hbar)$ the anisotropy field, $\gamma_{\mathbf{k}} = z_0^{-1}\sum_j e^{i\mathbf{k}\cdot\boldsymbol{\delta}_j}$ the lattice structure factor, and $z_0$ the coordination number. To diagonalize Eq.~\eqref{eq:HAFM}, we introduce the Bogoliubov transformation
\begin{equation}
a_\vk = u_\vk\, \alpha_\vk - v_\vk\, \beta^\dag_{-\vk}\,, \quad
b^\dag_{-\vk} = -v_\vk\, \alpha_\vk + u_\vk\, \beta^\dag_{-\vk}\,,
\end{equation}
where the coherence factors are given by
\begin{equation}
u(\mathbf{k}), v(\mathbf{k}) = \sqrt{\frac{\gamma D + \gamma A \pm \omega_0(\vk)}{2\,\omega_0(\vk)}}\,,
\end{equation}
with
\begin{equation}
\omega_0(\vk) = \gamma \sqrt{A^2+2AD + \frac{1}{3}D^2 a^2 k^2}.
\label{eq:uv_afm}
\end{equation}
where $a$ is the lattice constant.
In this basis, Eq.~\eqref{eq:HAFM} takes the diagonal form
\beq
\mathcal{H}^{\text{\tiny(AFM)}}_{\text{m}}=\hbar\sum_{\vk} \left[ \om_{\text{m},\al} (\vk) \al_{\vk}^\dag \al_{\vk} + \om_{\text{m},\be} (\vk) \be_{\vk}^\dag\be_{\vk}\right]\,,
\label{eq:AFM}
\eeq
where
\beq
\omega_{\text{m},\alpha(\beta)}(\vk) = \omega_0(\vk) \pm \gamma B_0\,.
\label{eq:AFMshift}
\eeq
Equation~\eqref{eq:AFMshift} shows that the two magnon branches experience opposite frequency shifts \(\pm \gamma B_0\) in response to the external field, reflecting their opposite chirality. As illustrated in Fig.~\ref{fig:schematic1}, the AFM mode \(\alpha\) (\(\beta\)) corresponds to a collective precession of both sublattices in a counterclockwise (clockwise) sense at frequency \(\omega_{m,\alpha (\beta)}\). The dynamical asymmetry between the two magnon branches is directly reflected in their angular–momentum content. Namely, \(\alpha^\dagger_{\mathbf{k}}\) and \(\beta^\dagger_{\mathbf{k}}\) create magnons carrying angular momentum \(-\hbar\) and \(+\hbar\), respectively, as it can be seen clearly by rewriting the total $z$-component of the angular momentum as
\begin{equation}
S^{z} =  \hbar \sum_{i} \left( S_{A,i}^z+ S_{B,i}^z\right) = -\hbar  \sum_{\mathbf{k}} \left( \alpha^\dagger_{\mathbf{k}} \alpha_{\mathbf{k}} - \beta^\dagger_{\mathbf{k}} \beta_{\mathbf{k}} \right)\,.
\label{eq:Stot}
\end{equation}
On the other hand,  by inserting Eq.~\eqref{eq:hp1} into Eq.~\eqref{eq:Hm} for \( \mathcal{J} > 0 \), and performing a few straightforward algebraic manipulations, we obtain the FM  Hamiltonian as
\begin{equation}
\mathcal{H}_{\text{m}} =  \sum_{\mathbf{k}} \hbar \omega_{\text{m}}(\mathbf{k})\, a^\dagger_{\mathbf{k}} a_{\mathbf{k}}\,,
\end{equation}
with the single–branch magnon dispersion  given by
\begin{equation}
\omega_{\text{m}}(\mathbf{k}) = \gamma \left( \frac{1}{6}D a^2k^2 + A + B_0 \right).
\end{equation}
The presence of only a single precession chirality in the FM case, in contrast to the two oppositely polarized magnon modes in the AFM, gives rise to a distinct angular momentum structure. In the next section, we show how this distinction constrains the symmetry–allowed channels through which angular momentum can be transferred to the lattice.

\subsection{Magnon–phonon coupling }\label{sec:mec}

In standard magnetoelastic theory~\cite{birss1966symmetry,callen1963static,brown1965theory}, the spin–lattice coupling is introduced phenomenologically by considering the most general set of symmetry-allowed bilinear couplings between strain and components of the magnetization, with coefficients that quantify the strength of each term.  While this approach correctly captures the tensor structure enforced by crystal symmetries, it leaves both the microscopic origin and the magnitude of the coupling parameters unspecified.

Recent works~\cite{mankovsky2022angular,weissenhofer2023rotationally,ruckriegel2020angular} have introduced an alternative, microscopic route that relies on explicitly restoring the rotational invariance of the spin Hamiltonian. To appreciate this construction, note that in Eq.~(\ref{eq:Hm}) the exchange and Zeeman terms are rotationally invariant, whereas the single–ion anisotropy term 
breaks global rotational symmetry by selecting a fixed spatial axis $\hat{\textbf{z}}$.
To recover a rotationally invariant description, one must promote the fixed anisotropy axis 
$\hat{\textbf{z}}$
to a local unit vector $\hat{\textbf{e}}_z$ that follows the instantaneous orientation of the crystalline environment and  depends explicitly on the lattice displacement field $\mathbf{u}(\mathbf{r})$. As illustrated in Fig.~\ref{fig:schematic2}, an atomic position $\mathbf{R}_i$ with nearest neighbors along the equilibrium $\hat{\mathbf{z}}$ axis located at $\mathbf{R}_{i,z^\pm}=\mathbf{R}_i\pm a\,\hat{\mathbf{z}}$ are displaced to  $\mathbf{r}_i = \mathbf{R}_i + \mathbf{u}_i$ and $\mathbf{r}_{i,z^\pm}=\mathbf{R}_{i,z^\pm}+\mathbf{u}_{i,z^\pm}$, respectively. Defining the local easy–axis by the neighbor chord
\begin{equation}
\hat{\mathbf{e}}^\pm_z(\mathbf{r}_i)=
\frac{\mathbf{r}_{i,z^\pm}-\mathbf{r}_{i}}
{\big|\mathbf{r}_{i,z^\pm}-\mathbf{r}_{i}\big|}\,,
\label{eq:localaxis}
\end{equation}
naturally restores rotational invariance by allowing the easy–axis direction to co-rotate with the lattice.
The anisotropy term in Eq.~(\ref{eq:Hm}) therefore becomes
\begin{equation}
- K \sum_i (\mathbf{S}_i \cdot \hat{\textbf{z}})^2 \rightarrow - \frac{K}{2} \sum_i (\mathbf{S}_i \cdot \hat{\mathbf{e}}^+_z)^2 - \frac{K}{2} \sum_i (\mathbf{S}_i \cdot \hat{\mathbf{e}}^-_z)^2\,.
\label{newani}
\end{equation}
For small displacements, expanding Eq.~(\ref{eq:localaxis}) to linear order in the relative deformation $\Delta \mathbf{u}_i=\mathbf{u}_{i,z^+}-\mathbf{u}_{i,z^-}$ modifies the local anisotropy axis accordingly. Substituting this into Eq. \eqref{newani} yields
\begin{equation}
\begin{split}
   & -K\sum_i(\mathbf{S}_i \cdot \hat{\mathbf{z}})^2
    -\frac{K}{a}\sum_i
    (\mathbf{S}_i \cdot \hat{\mathbf{z}})
    \big[\mathbf{S}_i \cdot (\mathbf{u}_{i,z^+}-\mathbf{u}_{i,z^-})\big]\,.
    \label{eq:Hmec}
    \end{split}
\end{equation}
 In the continuum limit, the finite difference in Eq.~(\ref{eq:Hmec}) reduces to $(\mathbf{u}_{i,z^+}-\mathbf{u}_{i,z^-})/2a\!\to\!\partial_z\mathbf{u}(\mathbf{r})$, so that $\nabla\times\mathbf{u}$ acts as the local rotational field mediating the exchange. Accordingly, the second term in Eq.~\eqref{eq:Hmec} can be identified as the magnon–phonon coupling Hamiltonian
 \begin{equation}
     \mathcal{H}_{\text{mec}}=
    -\frac{K}{a}\sum_i
    (\mathbf{S}_i \cdot \hat{\mathbf{z}})
    \big[\mathbf{S}_i \cdot (\mathbf{u}_{i,z^+}-\mathbf{u}_{i,z^-})\big]\,,
     \label{MECHamilt}
 \end{equation}
 which  describes the transfer of angular momentum between the spin and lattice subsystems: a lattice rotation exerts a torque on the local spin, while precessing spins feed angular momentum back into the lattice. 
For the AFM case, we find that, to quadratic order in bosonic operators, Eq.~\eqref{MECHamilt} can be written explicitly as 
\begin{equation}
\label{eq:Hmec_AFM_full}
\mathcal{H}^{\text{\tiny (AFM)}}_{\text{mec}} = -K \sum_{\mathbf{k},\lambda} \eta_\lambda(\mathbf{k}) \left( \alpha_\mathbf{k}^\dagger - \beta_{-\mathbf{k}} \right) \left( c_{\mathbf{k},\lambda} + c_{-\mathbf{k},\lambda}^\dagger \right) + \text{H.c.},
\end{equation}
where $\lambda\in\{+,-,\parallel\}$ and 
 \begin{equation}
 \begin{split}
\eta_\lam(\mathbf{k}) &= -2S(u^*(\mathbf{k}) + v^*(\mathbf{k})) \sqrt{\frac{\hbar S}{4\Bar{\rho}V \omega_{\text{p},\lambda}(k)}}k e^{i \phi_\mathbf{k}} \\
&\times [i\delta_{+,\lambda} (\cos\theta_\mathbf{k}+1) + i\delta_{-,\lambda} (\cos\theta_\mathbf{k}-1) - \delta_{\parallel,\lambda} \sin\theta_\mathbf{k}]\,.
\end{split}
\label{eq:eta}
\end{equation}
Similarly, in the FM case, where the magnon spectrum consists of a single branch, the magnetoelastic coupling reduces to
\begin{equation}
\label{eq:Hmec_FM_full}
\mathcal{H}^{\text{\tiny (FM)}}_{\text{mec}} = -K \sum_{\mathbf{k},\lambda} \eta_\lambda(\mathbf{k}) a_\mathbf{k}^\dagger \left( c_{\mathbf{k},\lambda} + c_{-\mathbf{k},\lambda}^\dagger \right) + \text{H.c.},
\end{equation}
where $\eta_\lambda(\mathbf{k})$ is given by Eq.~\eqref{eq:eta} with $u(\mathbf{k})=1$ and $v(\mathbf{k})=0$ for all $\mathbf{k}$. For concreteness, in the following, we focus on propagation along the anisotropy axis, \(\theta_{\mathbf{k}} = 0\) or \(\pi\), for which the magnetoelastic vertex does not couple to the LA branch. In this geometry, rotations about \(\hat{\mathbf{z}}\) simply relabel the transverse polarizations: without loss of generality, we set \(\phi_{\mathbf{k}} = 0\).

It is instructive to note that both Eqs.~\eqref{eq:Hmec_AFM_full} and~\eqref{eq:Hmec_FM_full} contain only operators that preserve the axial component of the total angular momentum. This structure follows from rotational invariance: since the full Hamiltonian commutes with \(J^{z} = S^{z} + L^{z}\), any decomposition of the magnetoelastic coupling into eigen-operators of the adjoint action of \(J^{z}\) can have support only in the sector with \(\Delta J^{z} = 0\).  As a result,  co-rotating  terms of the type \(\alpha_{\mathbf{k}}^{\dagger}c_{\mathbf{k},+}\) and \(\beta_{\mathbf{k}}^{\dagger}c_{\mathbf{k},-}\) are symmetry allowed, i.e.,
\begin{equation}
[J^z,\alpha_{\mathbf{k}}^{\dagger}c_{\mathbf{k},+}]
=
[J^z,\beta_{\mathbf{k}}^{\dagger}c_{\mathbf{k},-}]
=0\,,
\end{equation}
and, by the same reasoning, the co-rotating pair–creation and pair–annihilation terms
\(\alpha_{\mathbf{k}}^{\dagger}c_{-\mathbf{k},-}^{\dagger}\) and \(\beta_{\mathbf{k}}^{\dagger}c_{-\mathbf{k},+}^{\dagger}\) (together with their Hermitian conjugates) are also compatible with the symmetry, since each bilinear changes the spin and lattice contributions to \(J^{z}\) by equal and opposite amounts, leaving \(\Delta J^{z}=0\). 
On the other hand,  opposite–helicity pairs shift the angular momentum by two quanta, i.e., 
\begin{equation}
\begin{split}
[J^z,\alpha_{\mathbf{k}}^{\dagger}c_{\mathbf{k},-}]
=&-2\hbar\,\alpha_{\mathbf{k}}^{\dagger}c_{\mathbf{k},-}\neq 0\,,\\
[J^z,\beta_{\mathbf{k}}^{\dagger}c_{\mathbf{k},+}]
=&+2\hbar\,\beta_{\mathbf{k}}^{\dagger}c_{\mathbf{k},+}\neq 0\,,
\end{split}
\end{equation}
and, thus, cannot appear with finite weight in the  rotationally invariant magnetoelastic Hamiltonian~\eqref{eq:Hmec_AFM_full}.

Equations~\eqref{eq:Hmec_AFM_full} and~\eqref{eq:Hmec_FM_full} contain both
stationary and rapidly oscillating (counter-rotating) contributions.  To extract the physically relevant, near–resonant interactions -- those that remain stationary on timescales longer than the inverse hybridization gap -- we
move to the interaction picture with respect to the free magnon and phonon
Hamiltonians. In this frame, the operators evolve as  
\begin{equation}
\alpha_{\mathbf{k}}(t) = \alpha_{\mathbf{k}}\,e^{-i\omega_{\alpha}(\mathbf{k}) t}\,
\, \, \,
c_{\mathbf{k},\lambda}(t) = c_{\mathbf{k},\lambda}\,e^{-i\omega_{p,\lambda}(\mathbf{k}) t}\,,
\end{equation}
and analogously for \(\beta_{\mathbf{k}}(t)\) and \(a_{\mathbf{k}}(t)\).
Terms such as \( \alpha_{\mathbf{k}}^{\dagger} c_{-\mathbf{k},\lambda}^{\dagger} \) or 
\( \alpha_{\mathbf{k}} c_{-\mathbf{k},\lambda} \) oscillate at frequencies 
\(\sim \omega_{\alpha}(\mathbf{k}) + \omega_{p,\lambda}(\mathbf{k})\) and therefore 
average to zero over experimentally relevant timescales. 
By retaining only the slowly varying (energy–conserving) contributions, Eq.~\eqref{eq:Hmec_AFM_full} can be written as
\begin{equation}
\label{eq:Hmec_AFM_RWA}
\mathcal{H}^{\text{\tiny (AFM)}}_{\text{mec}} = -K \sum_\mathbf{k} \left[ \eta_+(\mathbf{k}) \alpha_\mathbf{k}^\dagger c_{\mathbf{k},+} \!-\! \eta_-(\mathbf{k}) \beta^\dagger_{\mathbf{k}} c_{\mathbf{k},-} \right] + \text{H.c.},
\end{equation}
which makes the chiral selectivity of the coupling transparent:
the right–handed magnon $\alpha_{\mathbf{k}}$, carrying angular momentum $-\hbar$, hybridizes efficiently only with the
co-rotating phonon $c_{\mathbf{k},+}$, while the left–handed magnon
$\beta_{\mathbf{k}}$, carrying angular momentum $\hbar$, hybridizes only with $c_{\mathbf{k},-}$.  

Each helicity therefore defines an independent dynamical sector, i.e., the coupled Hamiltonian
\(
\mathcal{H}^{\text{\tiny(AFM)}}=\mathcal{H}^{\text{\tiny(AFM)}}_m+
\mathcal{H}^{\text{\tiny(AFM)}}_p+
\mathcal{H}^{\text{\tiny(AFM)}}_{\text{mec}}
\)
is block–diagonal in the chiral basis and decomposes into two independent
\(2\times 2\) blocks. For each wave vector \(\mathbf{k}\), the block with helicity
\(\pm\) takes the form
\begin{equation}
\mathcal{H}_{\mathbf{k},\pm}=
\begin{pmatrix}
\hbar \omega_{\text{m},\alpha(\beta)}(\mathbf{k}) & -K\,\eta_{\pm}(\mathbf{k})\\[3pt]
-K\,\eta_{\pm}^{*}(\mathbf{k}) & \hbar \omega_{\text{p},\pm}(\mathbf{k})
\end{pmatrix}\,.
\end{equation}
An entirely analogous structure emerges in the ferromagnetic case. Here the magnon sector supports a single precession chirality: the mode $a_{\mathbf{k}}$ carries angular momentum $-\hbar$ and thus can exchange angular momentum only with the phonon branch of matching handedness. Equation
~\eqref{eq:Hmec_FM_full} can be rewritten as 

\begin{equation}
\label{eq:Hmec_FM_RWA}
\mathcal{H}^{\text{\tiny (FM)}}_{\text{mec}} = -K \sum_\mathbf{k} \eta_+(\mathbf{k}) a_\mathbf{k}^\dagger c_{\mathbf{k},+} + \text{H.c.}
\end{equation}
and the coupled
quadratic Hamiltonian \( \mathcal{H}^{\text{\tiny(FM)}} = \mathcal{H}^{\text{\tiny(FM)}}_m + \mathcal{H}^{\text{\tiny(FM)}}_p + \mathcal{H}^{\text{\tiny(FM)}}_{\text{mec}} \) in the co-rotating sector reduces to
\begin{equation}
\mathcal{H}_{\mathbf{k}}^{\text{\tiny(FM)}}
=
\begin{pmatrix}
\hbar \omega_\text{m}(\mathbf{k}) & -K\,\eta_{+}(\mathbf{k})\\[3pt]
-K\,\eta_{+}^{*}(\mathbf{k}) & \hbar \omega_{\text{p},+}(\mathbf{k})
\end{pmatrix}.
\label{eq:Hk_FM}
\end{equation}

\section{Magnon–Polaron modes}\label{sec:magpol1}

In this Section, we introduce the magnon–polaron normal modes that diagonalize the coupled spin–lattice Hamiltonians derived in Sec.~\ref{sec:mec}. For the FM system, the dynamics at fixed $\mathbf{k}$ are governed by the $2\times2$ matrix $\mathcal{H}^{\text{\tiny(FM)}}_{\mathbf{k}}$~\eqref{eq:Hk_FM}, whose diagonalization defines the hybrid eigenmodes through
\begin{equation}
\mathcal{H}^{\text{\tiny(FM)}}_{\mathbf{k}} |w_{\mathbf{k},i}\rangle
= \hbar \Omega^{\text{\tiny(FM)}}_{\mathbf{k},i} |w_{\mathbf{k},i}\rangle,
\qquad
|w_{\mathbf{k},i}\rangle =
\begin{pmatrix}
U^{\text{(m)}}_{\mathbf{k},i}\\[2pt]
U^{\text{(p)}}_{\mathbf{k},i}
\end{pmatrix},
\label{eq:eigvec}
\end{equation}
with $i=1,2$ labeling, respectively, the upper and lower magnon–polaron branches. The corresponding eigenfrequencies follow directly from the characteristic equation as
\begin{equation}
\begin{split}
\Omega_{\mathbf{k},i}^{\text{\tiny (FM)}} &=  \frac{1}{2} \left[\omega_\text{m}(\mathbf{k}) + \omega_{\text{p},+}(\mathbf{k}) \right]\\
&\pm \frac{\sqrt{\left[\omega_\text{m}(\mathbf{k}) - \omega_{\text{p},+}(\mathbf{k})\right]^2 + 4K^2|\eta_{+}(\mathbf{k})|^2/\hbar^2}}{2}.
\end{split}
\label{eq:mpfreqFM}
\end{equation}
The normalized eigenvector $|w_{\mathbf{k},i}\rangle$ specifies the magnon and phonon content of each hybrid mode. In second–quantized form, the corresponding magnon–polaron creation operator is
\begin{equation} \Gamma^\dag_{\mathbf{k},i} = U^{\text{(m)}}_{\mathbf{k},i}a^\dag_\mathbf{k} + U^{\text{(p)}}_{\mathbf{k},i}c^\dag_{\mathbf{k},+}, \label{eq:mpFM} \end{equation} which defines the canonical transformation from bare magnons and transverse phonons to the FM magnon–polaron basis.  The magnonic weight $s_{\mathbf{k},i} = |U^{\text{(m)}}_{\mathbf{k},i}|^2$ of the $i$-th band can be written as 
\begin{align}
s_{\mathbf{k},i} = \frac{1}{2} + \frac{\omega_{\text{m}}(\mathbf{k}) - \omega_{\text{p},+}(\mathbf{k})} {2\bigl(2\Omega^{\text{\tiny(FM)}}_{\mathbf{k},i} - \omega_{\text{m}}(\mathbf{k}) - \omega_{\text{p},+}(\mathbf{k})\bigr)} \,, 
\label{eq:magweight}
\end{align}
with $s_{\mathbf{k},i}\in[0,1]$. In terms of Eq.~\eqref{eq:magweight}, the total spin angular momentum of the FM system takes the compact form 
\begin{equation} S^{z} = -\hbar\sum_{\mathbf{k}} \bigl( s_{\mathbf{k},1}\,\Gamma^{\dagger}_{\mathbf{k},1}\Gamma_{\mathbf{k},1} + s_{\mathbf{k},2}\,\Gamma^{\dagger}_{\mathbf{k},2}\Gamma_{\mathbf{k},2} \bigr)\,. \label{eq:SzFM} \end{equation}

\begin{table}[h]
    \centering
     \caption{Selected YIG  and MnF$_2$ parameters \cite{flebus17,rezende19,Melcher70}.}
     \label{tab:my_label}
     \setlength{\tabcolsep}{0.17cm} 
    \begin{tabular}{lcccc}
    \hline\hline
      & Symbol & YIG & MnF$_2$ & Unit \\ \hline
      Spin & $S$ & 20 & 5/2& -\\
      Exchange constant & $\mathcal{J}$ & $0.05$ & -$0.203$ &  meV \\ 
      Anisotropy strength & $K$ &  $0.002$ & 0.019 &  meV \\
      External field & $B_0$ & 3.96  & 1 & T\\ 
      Spin-flop field & $B_\text{SF}$ & - & 9.3  & T\\
      TA Phonon speed & $v_{\pm}$ & $3900$ & $2800$ & m/s\\ 
      Lattice parameter & $a/c$ &$12.36$  & 3.3 & \AA \\
      Mass density &  $\bar{\rho}$& $5170$ & $3980$ & kg/m$^3$ \\
      MEC Enhancement & - & 30 & 5 & -\\   
        \hline\hline
    \end{tabular}
\end{table}

\begin{figure}[htb]
\centering
\ig[width=0.48 \textwidth, trim={1.4cm 0.3cm 3cm 1.2cm}, clip]{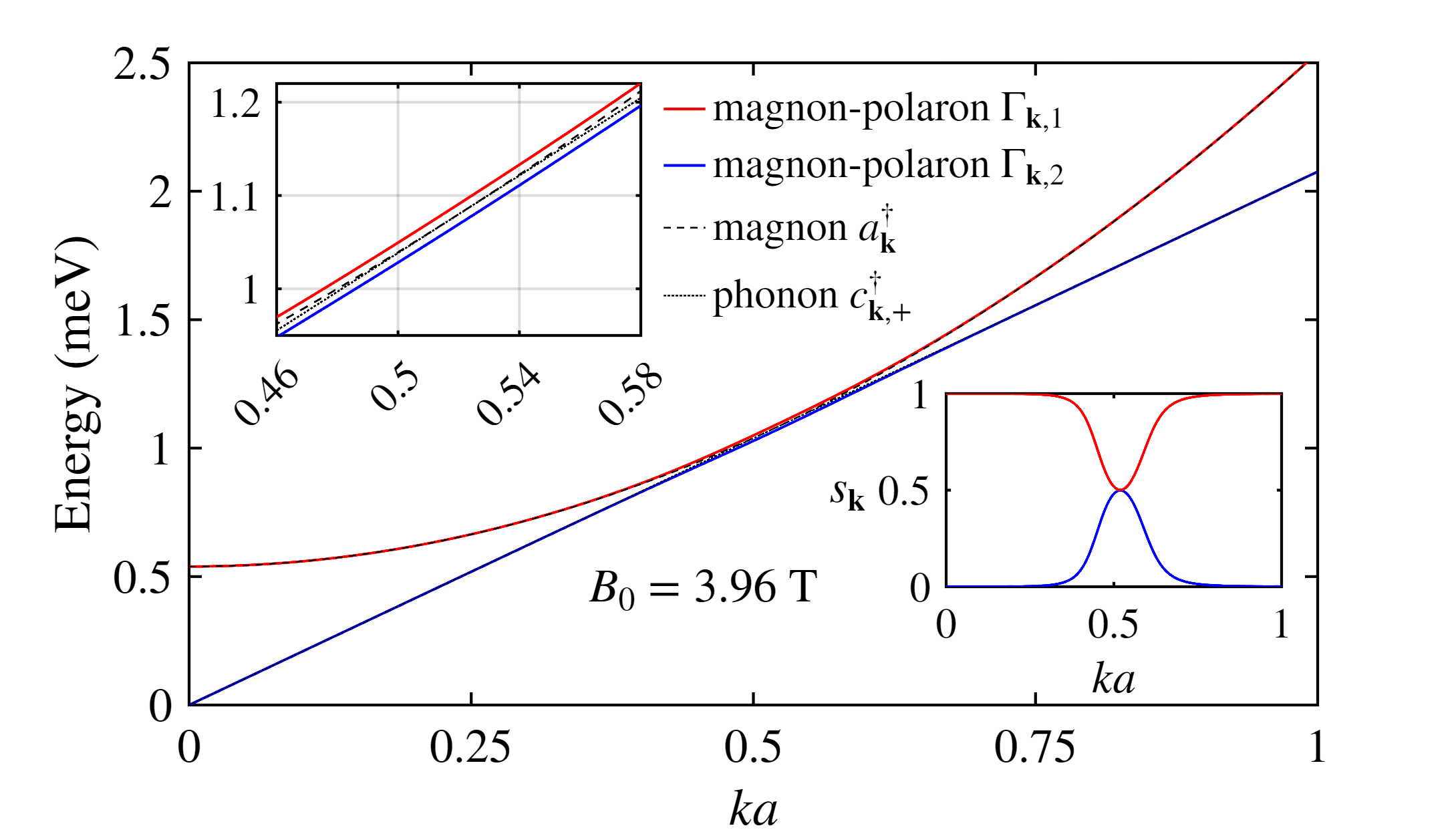}
\caption{Ferromagnetic magnon (dashed), transverse acoustic phonon (dotted), and magnon–polaron (red and blue) dispersions for 
$\mathbf{k} \parallel \hat{\mathbf{z}}$ and $\mathbf{B}_0 \parallel \hat{\mathbf{z}}$ computed using material parameters for YIG, see Table \ref{tab:my_label}. For clarity, the magnetoelastic coupling constant $\eta_+$ has been enhanced 30–fold so that the anti-crossing is visible on the scale of the plot. Hybridization between the magnon and the circularly polarized TA phonon with matching handedness produces two magnon–polaron branches with mixed angular–momentum character, determined by their respective magnon weights $s_\mathbf{k}$. For $B_0=3.96 \,\text{T}$, the TA branch becomes locally tangent to the magnon branch, maximizing the phase space for magnon–polaron formation and yielding $s_\mathbf{k}\simeq 1/2$ for both modes over a broad range of wave vectors (see insets). }
\label{fig:fm_plt}
\end{figure}

By construction, the lower branch carries the complementary weight \(s_{\mathbf{k},2} =1 - s_{\mathbf{k},1} \), ensuring that the total angular momentum is conserved. As shown by Eq.~(\ref{eq:SzFM}), the angular momentum is redistributed between the two hybrid modes in proportion to their magnonic content, and their combined contribution matches the spin angular momentum of the uncoupled magnon system. If the phonons themselves carried intrinsic angular momentum, that contribution would likewise be shared between the hybrid modes. In the present case, however, the lattice is centrosymmetric and phonons are not coupled to any TRS–breaking degrees of freedom other than the magnetic order: any lattice angular momentum arises solely through hybridization with the spin sector.  Taking YIG as a representative material, we plot in Fig.~\ref{fig:fm_plt} the FM magnon–polaron dispersions and their angular–momentum content at the magnetic field for which the uncoupled magnon and TA phonon dispersions become locally tangent. As shown in Ref.~\cite{flebus17}, this ``tangent condition'' maximizes the phase space volume for hybridization: the uncoupled dispersions share both their energy and group velocity at a single wavevector, which causes them to touch rather than cross. Under these conditions, $s_{\mathbf{k},1}$ and $s_{\mathbf{k},2}$ approach comparable values over the broadest range of momenta, providing an upper bound within our model for the transfer of spin angular momentum from the upper to the lower branch.

Following the same procedure for the AFM case, we find the magnon–polaron eigenfrequencies as  
\begin{equation}
\begin{split}
\Omega_{\mathbf{k},i}^{\text{\tiny(AFM)}}
 &= \frac{1}{2}\left[\omega_{\text{m},\lambda}(\mathbf{k})
 + \omega_{\text{p},\lambda}(\mathbf{k}) \right] \\
 &\pm\frac{\sqrt{\left[\omega_{\text{m},\lambda}(\mathbf{k})
 - \omega_{\text{p},\lambda}(\mathbf{k})\right]^2
 + 4K^2|\eta_{\lambda}(\mathbf{k})|^2/\hbar^2}}{2}\,,
 \label{eq:AFMdisp}
 \end{split}
 \end{equation}
 where $\lambda$ takes the values $(\alpha,+)$ and $(\beta,-)$, labeling the two chiral magnetoelastic channels selected by Eqs.~\eqref{eq:Hmec_AFM_RWA}: the right–handed AFM magnon $\alpha_{\mathbf{k}}$ couples only to the $+$ phonon branch, while the left–handed magnon $\beta_{\mathbf{k}}$ couples only to the $-$ branch. For each channel, Eq.~\eqref{eq:AFMdisp} yields an upper  and lower magnon–polaron branch: 
 we collect the resulting four AFM magnon–polaron modes into a single band index $i=1,\dots,4$, with $i=1,2$ corresponding to the upper and lower branches of $(\alpha,+)$ and $i=3,4$ to those of $(\beta,-)$. As in the FM case, the magnonic weight $s_{\mathbf{k},i} = |U^{(m)}_{\mathbf{k},i}|^2$ determines how the conserved axial angular momentum is distributed among the hybrid branches. Expressed in the magnon–polaron basis, the magnon angular momentum~\eqref{eq:Stot} takes the form
 \begin{align}
    S^{z} = &-\hbar \sum_{\mathbf{k}} \big[  s_{\mathbf{k},1} \Gamma^\dag_{\mathbf{k},1} \Gamma_{\mathbf{k},1} + (1 - s_{\mathbf{k},1}) \Gamma^\dag_{\mathbf{k},2} \Gamma_{\mathbf{k},2} \notag \\
    & - s_{\mathbf{k},3} \Gamma^\dag_{\mathbf{k},3} \Gamma_{\mathbf{k},3} - (1-s_{\mathbf{k},3}) \Gamma^\dag_{\mathbf{k},4} \Gamma_{\mathbf{k},4} \big]\,,
    \label{eq:mpAngAFM}
\end{align}

where $s_{\mathbf{k},1}$ and $1-s_{\mathbf{k},1}$ are the magnonic weights of the right–handed $(\alpha,+)$ upper and lower branches, while $s_{\mathbf{k},3}$ and $1-s_{\mathbf{k},3}$ are those of the left–handed $(\beta,-)$ upper and lower branches. Equation~\eqref{eq:mpAngAFM} makes explicit that angular–momentum exchange takes place independently within each chiral sector: the $(\alpha,+)$ hybrids contribute to the $-\,\hbar$ channel, and the $(\beta,-)$ hybrids to the $+\,\hbar$ channel.
\begin{figure}[htb]
\centering
\sfig[]{\ig[width=0.48 \textwidth, trim={1.4cm 0.3cm 3cm 1.2cm}, clip]{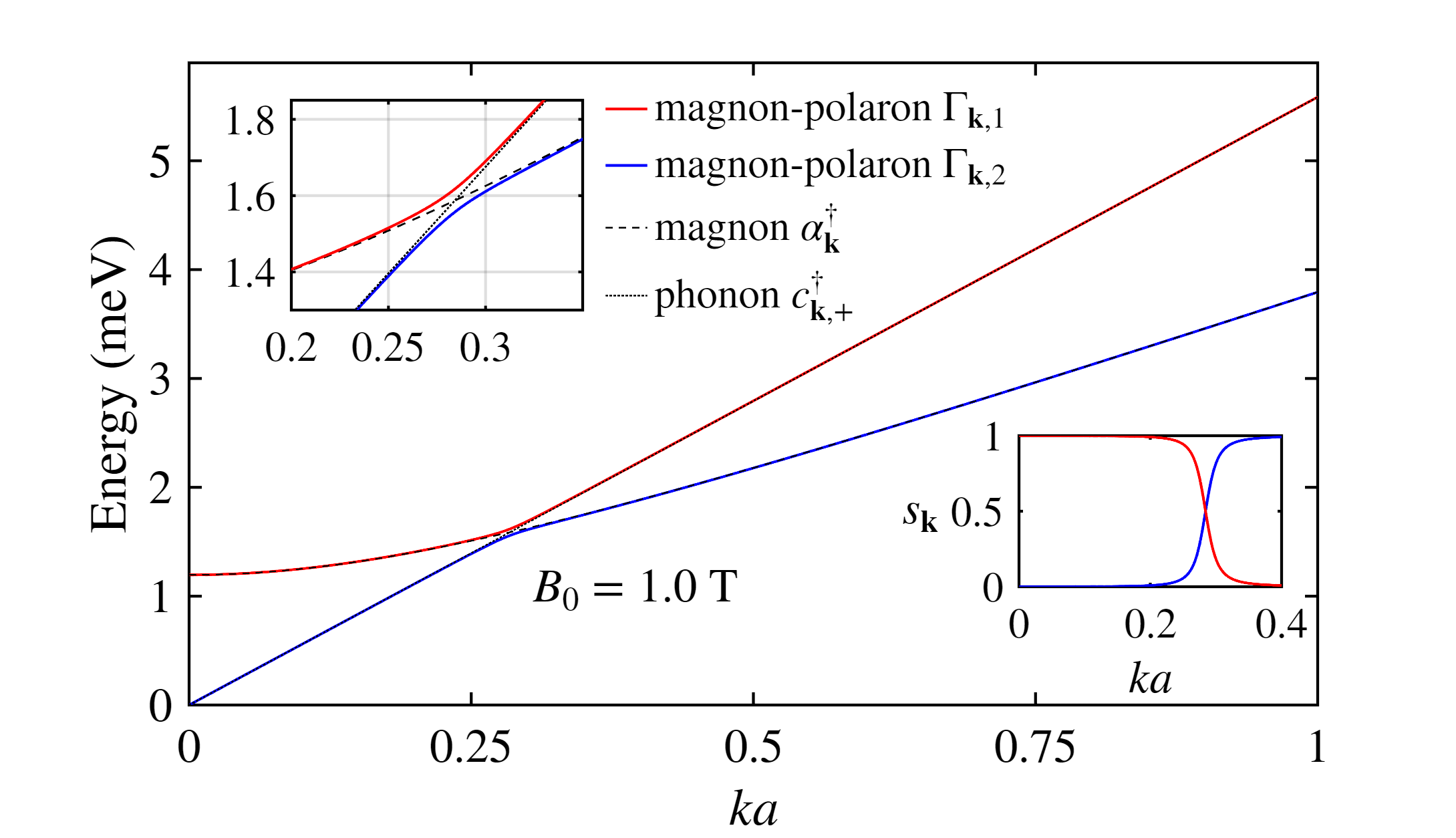}\label{fig:afm_plta}} \\ 
\sfig[]{\ig[width=0.48 \textwidth, trim={1.4cm 0.3cm 3cm 1.2cm}, clip]{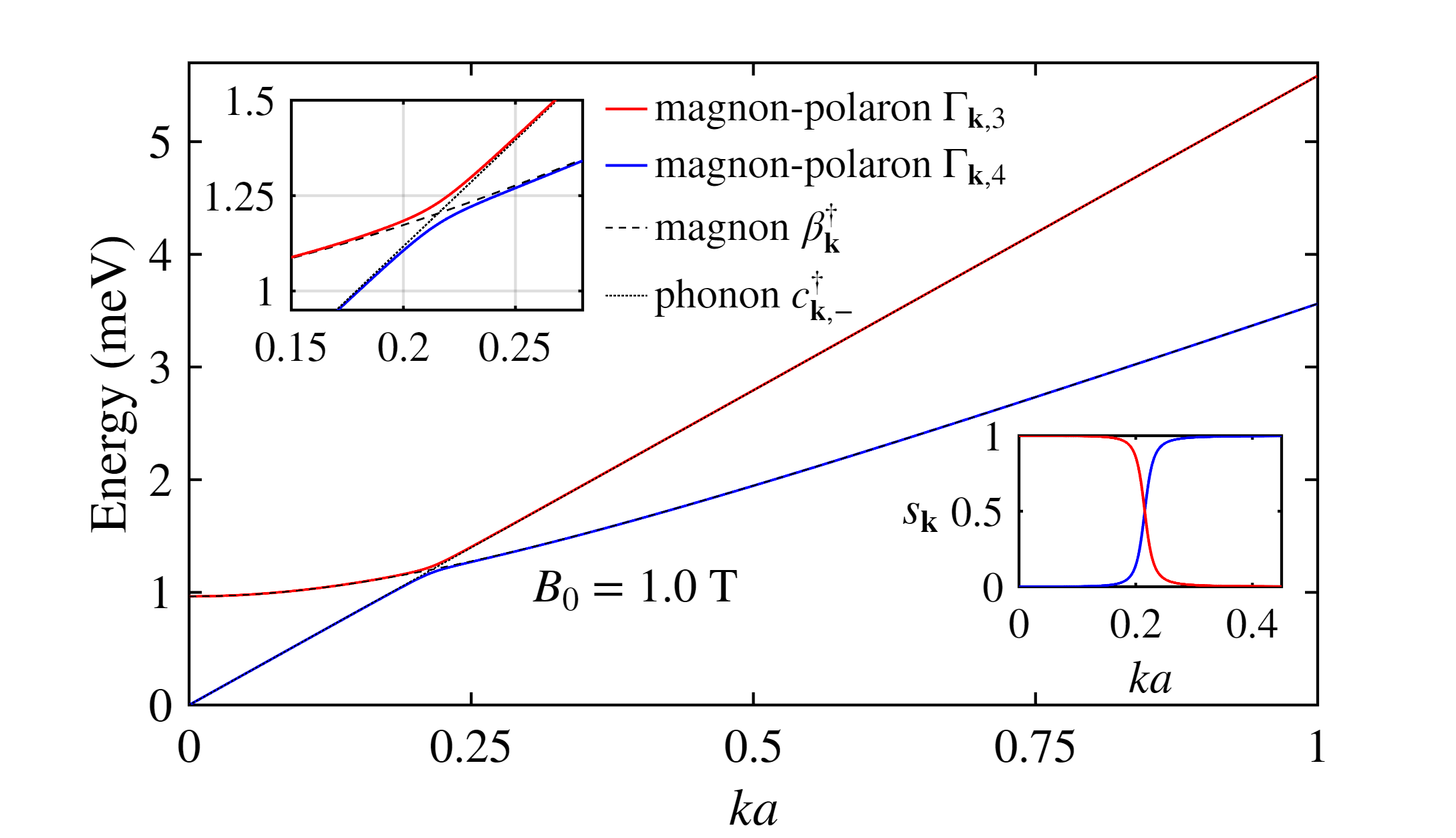}\label{fig:afm_pltb}}
\caption{Antiferromagnetic magnons (dashed), transverse acoustic phonons (dotted), and magnon–polaron (red and blue) dispersions for 
$\mathbf{k} \parallel \hat{\mathbf{z}}$ and $\mathbf{B}_0 \parallel \hat{\mathbf{z}}$ computed using material parameters for MnF$_2$, see Table \ref{tab:my_label}. For clarity, the magnetoelastic coupling constants $\eta_\pm$ have been enhanced 5–fold so that the anti-crossings are visible on the scale of the plot. Hybridization between the magnons and the circularly polarized TA phonons with matching handedness produces two pairs of magnon–polaron branches, one pair in each chiral channel, with mixed angular–momentum character determined by their respective magnon weights $s_\mathbf{k}$ (see insets). 
(a) Right–handed channel.
(b) Left–handed channel.}
\end{figure}

Taking MnF$_2$ as an illustrative material, Figs.~\ref{fig:afm_plta} and \ref{fig:afm_pltb}  display, respectively, the dispersions of the right– and left–handed antiferromagnetic magnon–polaron branches and their associated angular–momentum content. Because the two magnon helicities experience opposite Zeeman shifts, a finite field brings one chiral sector closer to resonance with the phonons and enhances its hybridization, while the opposite sector remains largely off–resonant. In the near–resonant sector, the avoided crossing and the smooth interchange of $J^z$ between the upper and lower branches are clearly visible; in the more detuned sector, the increasing field shifts the magnon and phonon dispersions apart, reducing their hybrid character and suppressing angular–momentum transfer over a broader range of $\mathbf{k}$.

\section{Magnon–Polaron Transport}\label{sec:transport}

In this Section, we develop a linear–response transport theory for magnon–polaron quasiparticles in FMs and AFMs, focusing on spin and heat currents driven by a bulk temperature gradient. We first derive the chemical potential of the hybrid excitations, showing that a single thermodynamic parameter governs both upper and lower branches, with a sign structure that distinguishes the FM and AFM cases. We then formulate Boltzmann transport within the relaxation–time approximation, obtain explicit expressions for the spin and heat currents, and linearize the kinetic equations to extract the corresponding Onsager coefficients.

\subsection{Chemical Potential}\label{sec:transport_chem}

Previous works have shown that introducing a magnon chemical potential \(\mu\) is essential for describing spin and heat transport in low–damping ferromagnetic and antiferromagnetic insulators~\cite{Cornelissen2016Magnon,Demidov2017Chemical,du17,hoogeboom2020nonlocal,olsson2020pure}.
 In these frameworks, \(\mu\) is the chemical potential of a magnon gas, defined as the thermodynamic variable conjugate to the total angular momentum \(S^{z}\) evaluated with respect to a fixed global spin axis \(\hat{\mathbf{z}}\). 
 Here we move beyond this purely magnonic picture and enter the hybridized regime: we treat the magnon–polaron modes as the relevant bosonic carriers and introduce a single chemical potential, again denoted \(\mu\), as the thermodynamic variable conjugate to the conserved total angular momentum \(J^{z}\) enforced by the rotationally invariant spin–lattice coupling.

We begin by considering the FM case,  where both hybrid branches reside in the $-\hbar$ chiral sector and exchange angular momentum only with one another, as shown in Fig. \ref{fig:fm_plt}. 
As discussed in Sec.~\ref{sec:magpol1}, in our model the conserved axial angular momentum resides entirely in the magnon sector and, after hybridization, is distributed among the magnon–polaron branches according to their magnonic weights $s_{\mathbf{k},i}$, whereas the energy receives contributions from both spin and lattice through the hybrid dispersions. Thus we can write
\begin{equation}
J^{z} \;=\; -\hbar \sum_{i=1}^{2}\sum_{\mathbf{k}} s_{\mathbf{k},i}\, n_{\mathbf{k},i},
\qquad
U \;=\; \hbar \sum_{i=1}^{2}\sum_{\mathbf{k}} \Omega_{\mathbf{k},i}^{\text{\tiny (FM)}}\, n_{\mathbf{k},i} .
\label{eq:FM_Jz_U}
\end{equation}
where $n_{\mathbf{k},i}$ is the occupation number of mode $\mathbf{k}$ in the $i$-th magnon–polaron branch.
To determine the equilibrium mode occupations, we maximize the total number of microstates 
$\ln \tilde{\Omega}$ 
subject to the conservation of $J^z$ and $U$~\eqref{eq:FM_Jz_U}. 
For a bosonic subsystem with degeneracy $g_{\mathbf{k},i}$ and occupation $n_{\mathbf{k},i}$, the number of microstates is
\begin{equation}
    \tilde{\Omega}_{i}  \; = \prod_{\mathbf{k}}
    \frac{(n_{\mathbf{k},i} + g_{\mathbf{k},i} - 1)!}
         {n_{\mathbf{k},i}!\,(g_{\mathbf{k},i} - 1)!}\,.
    \label{eq:micro}
\end{equation}
The total number of configurations is 
$\tilde{\Omega}=\tilde{\Omega}_1\,\tilde{\Omega}_2$.
Using Stirling’s approximation $\ln N!\approx N\ln N - N$, we maximize the Lagrangian
$\ln\tilde{\Omega}-(\lambda/\hbar) J^z-\beta U$,
where $\lambda$ and $\beta$ are Lagrange multipliers enforcing
conservation of angular momentum and energy, respectively. Varying with respect to $n_{\mathbf{k},i}$ for each mode $(\mathbf{k},i)$ with degeneracy $g_{\mathbf{k},i}$ yields
\begin{equation}
\ln\!\left(\frac{n_{\mathbf{k},i}+g_{\mathbf{k},i}}{n_{\mathbf{k},i}}\right)
= \beta\,\hbar\,\Omega_{\mathbf{k},i}^{\text{\tiny (FM)}} - \lambda s_{\mathbf{k},i}\,.
\label{eq:lagrange-eq}
\end{equation}
Using Eq.~\eqref{eq:lagrange-eq}, the entropy variation of the $i$-th mode can be written as
\begin{align}
\mathrm{d}S_i
&= k_B \sum_{\mathbf{k}}
\ln\!\left(\frac{n_{\mathbf{k},i} + g_{\mathbf{k},i} - 1}{n_{\mathbf{k},i}}\right)\,
\mathrm{d}n_{\mathbf{k},i}
\notag\\
&= k_B \sum_{\mathbf{k}}
\big(\beta\,\hbar\,\Omega_{\mathbf{k},i}^{\text{\tiny (FM)}} - \lambda s_{\mathbf{k},i}\big)\,
\mathrm{d}n_{\mathbf{k},i}
\notag\\
&= \frac{1}{T}\,\mathrm{d}U_i \;+\; k_B (\lambda/\hbar) \mathrm{d}J_i^z,
\label{eq:Sb-variation-correct-1}
\end{align}
where $\beta=1/(k_BT)$ and the differential total energy and total angular momentum are $\mathrm{d}U_i=\hbar\sum_{\mathbf{k}}\Omega_{\mathbf{k},i}^{\text{\tiny (FM)}}\,\mathrm{d}n_{\mathbf{k},i}$ and
$\mathrm{d}J_i^z=-\hbar \sum_{\mathbf{k}} s_{\mathbf{k},i}\,\mathrm{d}n_{\mathbf{k},i}$, respectively.
From Eq.~\eqref{eq:Sb-variation-correct-1}, we identify the intensive variable
conjugate to the conserved angular momentum as
\begin{equation}
\mu \equiv \lambda k_B T\,.
\end{equation}
The hybrid modes couple to this global chemical potential with a weight equal to
their magnonic fraction $s_{\mathbf{k},i}$. Substituting this relation into
Eq.~\eqref{eq:lagrange-eq} yields, for $n_{\mathbf{k},i}\gg 1$, the
Bose--Einstein distributions:
\begin{equation}
f_{\mathbf{k},i}
= \frac{1}{\exp\!\big\{\beta\big(\hbar\,\Omega_{\mathbf{k},i}^{\text{\tiny (FM)}}-\mu\, s_{\mathbf{k},i}\big)\big\}-1}.
\label{eq:BE-weights}
\end{equation}
Equation~\eqref{eq:BE-weights} shows that both hybrid branches are governed by a single chemical potential \(\mu\), conjugate to the conserved \(J^{z}\), while each branch couples to \(\mu\) with a strength set by its magnonic weight \(s_{\mathbf{k},i}\). 

In the weak–coupling limit, where hybridization between magnons and phonons is negligible and the magnonic weight approaches \(s_{\mathbf{k},i} \rightarrow 1\,(0)\) for the upper (lower) branch, the corresponding occupation functions continuously reduce to those of a gas of nearly pure magnons and of effectively spinless phonons. In this regime, the upper branch inherits the full angular–momentum content and couples to the chemical potential in the same way as a conventional magnon mode, while the lower branch decouples from \(\mu\) and behaves as a purely thermal, spinless phonon bath. By contrast, nearby  the avoided crossing where \(s_{\mathbf{k},i} \simeq 1/2\), the two hybrid modes share both angular momentum and chemical potential on an equal footing.

For the AFM system, 
the four magnon–polaron bands \(\Gamma_i\) likewise organize into two chiral sectors: modes \(i=1,2\) belong to the \(-\hbar\) channel and modes \(i=3,4\) to the \(+\hbar\) channel. Introducing a chirality factor \(\nu_i = +1\) for \(i=1,2\) and \(\nu_i = -1\) for \(i=3,4\), the conserved axial angular momentum and the internal energy can be written compactly as
\begin{equation}
J^{z} \;=\; -\hbar \sum_{i=1}^{4}\sum_{\mathbf{k}} \nu_i\, s_{\mathbf{k},i}\, n_{\mathbf{k},i},
\quad
U \;=\; \hbar\sum_{i=1}^{4}\sum_{\mathbf{k}}\Omega_{\mathbf{k},i}^{\text{\tiny(AFM)}}\, n_{\mathbf{k},i} .
\label{eq:AFM_Jz_U}
\end{equation}
In an analogy with the FM case, maximizing the entropy with fixed $U$ and $J^z$ yields the Bose–Einstein distributions
\begin{equation}
f_{\mathbf{k},i}
= \frac{1}{\exp\!\big\{\beta\,\big(\hbar\,\Omega_{\mathbf{k},i}^{\text{\tiny(AFM)}} - \mu\,\nu_i\, s_{\mathbf{k},i}\,\big)\big\}-1},
\label{eq:AFM_BE}
\end{equation}
where $\mu$ is the chemical potential conjugate to $J^z$ associated with the AFM system. As with the FM system, each mode couples to the same $\mu$ with a weight proportional to its spin content $\nu_i s_{\mathbf{k},i}$. In the limit of vanishing magnetoelastic coupling, 
the magnonic weights approach $s_{\mathbf{k},1(3)} \to 1$ and
$s_{\mathbf{k},2(4)} \to 0$, restoring the two bare AFM magnon branches
$\alpha$ and $\beta$. 
Defining the band populations 
\( N_i = \sum_{\mathbf{k}} n_{\mathbf{k},i} \),
the decoupled total angular momentum simply becomes
\begin{equation}
J^z
= -\hbar \left(N_1 - N_3\right)\,,
\label{eq:limit}
\end{equation}
and the conjugate chemical potential reduces to the one introduced by Ref. \cite{flebus19}:
\begin{equation}
\mu
= \left.\frac{\partial U}{\partial (N_1 - N_3)}\right|_{S,V} \,.
\end{equation}

\subsection{Heat and Angular–Momentum Current}
We proceed to formulate a semiclassical description of bulk transport mediated by magnon–polaron quasiparticles. Our analysis rests on two key physical assumptions about the underlying magnon and phonon subsystems.
First, we assume rapid energy exchange between magnons and acoustic phonons, mediated by magnetoelastic scattering. When this rate exceeds the individual relaxation rates of the bare excitations, the two subsystems equilibrate locally, sharing a common temperature \(T(\mathbf{r})\).
Second, we require that magnon number be approximately conserved on the timescales relevant for transport. In low–damping magnetic insulators, spin-nonconserving processes---such as Gilbert damping and weak dipolar relaxation---enter at energy scales far below the exchange interaction that governs intraband thermalization. 
As a result, the hybrid magnon–polaron modes acquire a well–defined local chemical potential
\(\mu(\mathbf{r})\), which vanishes in equilibrium and becomes finite only in the presence of a small
thermodynamic drive, such as a temperature gradient.
 We consider transport along a generic direction $\hat{\mathbf{r}}$ and over a length $L$ that is taken to be much larger than both the thermal magnon and phonon de Broglie wavelengths and their respective mean free paths. 
Within this semiclassical picture, each magnon–polaron branch $i$ with wavevector $\mathbf{k}$ is described by a distribution function $f_{\mathbf{k},i}(\mathbf{r})$ whose dynamics obey the Boltzmann equation:
\begin{equation}
    \partial_t f_{\mathbf{k},i} 
    + \dot{\mathbf{r}}\cdot \partial_{\mathbf{r}} f_{\mathbf{k},i}
    + \dot{\mathbf{k}}\cdot \partial_{\mathbf{k}} f_{\mathbf{k},i}
    = \left.\frac{\partial f_{\mathbf{k},i}}{\partial t}\right|_{\mathrm{coll}}\,,
    \label{790}
\end{equation}
where $\dot{\mathbf{r}} = \partial_{\mathbf{k}}\Omega_{\mathbf{k},i}$ is the group velocity of the hybrid mode and $\dot{\mathbf{k}}$ encodes external forces. In the relaxation–time approximation, the collision integral becomes
\begin{equation}
    \left.\frac{\partial f_{\mathbf{k},i}}{\partial t}\right|_{\mathrm{coll}}
    = - \frac{\delta f_{\mathbf{k},i}}{\tau_{\mathbf{k},i}}\,,
    \label{795}
\end{equation}
where $\tau_{\mathbf{k},i}$ is the relaxation time and $\delta f_{\mathbf{k},i} = f_{\mathbf{k},i} - f^{(0)}_{\mathbf{k},i}$  the deviation from the local equilibrium distribution $ f^{(0)}_{\mathbf{k},i}$.
 In equilibrium, the magnon–polaron chemical potential vanishes since both magnon and phonon numbers are not conserved, implying that the equilibrium distribution of the $i$-th magnon–polaron branch is given by 
\begin{equation}
    f^0_{\mathbf{k},i}(\mathbf{r}) = \left[\text{exp}\left(\frac{\hbar \Omega_{\mathbf{k},i}}{k_B T}\right)-1\right]^{-1}\,.
    \label{eq:f0}
\end{equation}
Here, $T$ denotes the average temperature of the sample subjected to the local temperature profile
\begin{equation}
T(\mathbf{r}) = T + |\nabla T|\, \hat{\mathbf{r}} \cdot \mathbf{r}\,,
\end{equation}
where $|\nabla T|$ denotes a bulk temperature gradient set along the $\hat{\mathbf{r}}$ direction.  In the steady state and in the absence of external forces, the spatial variation of the local equilibrium distribution defines the angular–momentum, $ \mathbf{j}_{J^z}$,  and heat, $\mathbf{j}_q$, current densities as 
\begin{align}
     \mathbf{j}_{J^z} &= \int \frac{d^3\mathbf{k}}{(2\pi)^3} {\sum_i}\hbar\nu_i s_{\mathbf{k},i} (\partial_\mathbf{k} \Omega_{\mathbf{k},i})\delta f_{\mathbf{k},i},   \label{eq:currents2} \\
        \mathbf{j}_q &= \int \frac{d^3\mathbf{k}}{(2\pi)^3} \underset{i}{\sum} \hbar\Omega_{\mathbf{k},i} (\partial_\mathbf{k} \Omega_{\mathbf{k},i})\delta f_{\mathbf{k},i}\,, 
    \label{eq:currents1}
\end{align}
where $\nu_{i} $ is the chirality factor introduced for the AFM branches in Eq.~\eqref{eq:AFM_Jz_U}; in the FM case, both modes lie in the same chiral sector and thus $\nu_{1,2} = +1$. Equation~\eqref{eq:currents1} shows that, for a purely thermal drive, the total heat current $\mathbf{j}_q$ is independent of the magnonic content $s_{\mathbf{k},i}$ of the magnon–polaron modes, since both magnons and phonons contribute to energy transport irrespective of their relative weight in each branch.  By contrast, the angular–momentum current $\mathbf{j}_{J^z}$ is weighted by the magnonic fraction $s_{\mathbf{k},i}$ and the chirality factor $\nu_i$, and is therefore controlled by the “magnon–like’’ character of each branch. This structure is consistent with the phenomenological magnon–polaron transport theory of Ref.~\cite{flebus17}, where the spin current was obtained by projecting the hybrid modes onto the spin system and assuming that they inherit the magnon chemical potential; here, the same weighting by $s_{\mathbf{k},i}$ emerges directly from a microscopic, rotationally invariant formulation of magnetoelastic coupling.
In the limit of vanishing hybridization, i.e., $s_{\mathbf{k},i}\to 1$ ($0$) for purely magnonic (phononic) bands, Eq.~\eqref{eq:currents2} reduces to the standard expression for the magnonic spin current~\cite{Cornelissen2016Magnon}.


Within the linear response regime, i.e., $|\nabla T| L \ll T$, Eq. \eqref{eq:currents1} can be recast in matrix form as
\begin{equation}
\begin{pmatrix} \mathbf{j}_{J^z} \\ \mathbf{j}_{q} \end{pmatrix} \approx
-\begin{pmatrix} \boldsymbol\sigma &~&  \boldsymbol{\xi} \\  \boldsymbol\rho &~&  \boldsymbol{\kappa}\end{pmatrix} \begin{pmatrix} \nabla\mu \\  \nabla T \end{pmatrix}, 
\label{eq:currents3}
\end{equation}
where $\boldsymbol{\sigma}$ and $\boldsymbol{\kappa}$ denote, respectively, the angular–momentum and heat conductivity tensors, and $\boldsymbol{\xi}$ and $\boldsymbol{\rho}$ are the corresponding angular–momentum Seebeck and Peltier tensors.
For compactness, all entries of the response matrix in Eq.~\eqref{eq:currents3} can be written in terms of a set of generalized transport kernels:
\begin{align}
    L^{mn}_{\alpha \gamma} =& \beta \hbar \int \frac{d^3\mathbf{k}}{(2\pi)^3} \underset{i}{\sum} \tau_{\mathbf{k},i} (\nu_i s_{\mathbf{k},i})^m \frac{\partial \Omega_{\mathbf{k},i}}{\partial{\mathbf{k}_\alpha} } \frac{\partial \Omega_{\mathbf{k},i}}{\partial{\mathbf{k}_\ga}} \nonumber \\
    &\times \frac{ e^{\beta\hbar \Omega_{\mathbf{k},i}}  (\hbar\Omega_{\mathbf{k},i})^n}{(e^{\beta \hbar\Omega_{\mathbf{k},i}}-1)^2}\,,
    \label{eq:transport}
\end{align}
from which the transport coefficients follow as
\begin{align}
    \sigma_{\alpha \gamma} = L^{20}_{\alpha \gamma}\,,\quad
    \xi_{\alpha \gamma} = \frac{L^{11}_{\alpha \gamma}}{T}\,,\nonumber  \\
    \rho_{\alpha \gamma} = L^{11}_{\alpha \gamma}\,,\quad
    \kappa_{\alpha \gamma} = \frac{L^{02}_{\alpha \gamma}}{T}\,.
\end{align}

\section{Discussion and Conclusion}\label{sec:con}
In this work, we develop a microscopic framework that describes how magnetoelastic coupling shapes the thermodynamic and transport properties of magnon–polaron modes in collinear, $U(1)$-symmetric FM and AFM insulators. 
We focus on the regime in which magnon–phonon energy exchange is faster than quasiparticle–nonconserving processes, so that the axial component of the total angular momentum  can be treated as approximately conserved. In this setting, we show that global spin–rotation symmetry implies the existence of a single chemical potential, conjugate to the conserved quantity, that controls the nonequilibrium occupation of all the locally equilibrated magnon–polaron modes.

While our approach is general, we make the chiral structure of the hybrid modes explicit by focusing on the interaction between magnons and degenerate transverse acoustic branches that form circularly polarized phonons. In this representation, the chiral selectivity of the magnetoelastic interaction is most transparent: a magnon hybridizes only with the co-rotating circular phonon, yielding two similarly handed magnon–polaron branches in ferromagnets and two oppositely handed chiral sectors in collinear antiferromagnets. We find that, 
for ferromagnets, the two hybrid branches couple to this chemical potential with the same sign, weighted by their respective magnonic amplitudes. In collinear antiferromagnets, by contrast, the four hybrid branches group into two chiral sectors that carry opposite sign of angular momentum; each mode couples to the common chemical potential with a prefactor set by its magnonic content and its chirality.

Building on these insights, we derive a semiclassical transport theory for hybrid magnon–polaron carriers driven by temperature and chemical–potential gradients. The resulting structure is fully mode resolved: the heat current probes the entire magnetoelastic spectrum through the group velocities and lifetimes of all hybrid branches, whereas the angular–momentum current is weighted by the magnonic content and chirality of each mode. This procedure leads to compact expressions for the angular–momentum and heat currents that reduce exactly to the uncoupled magnon and phonon results when hybridization vanishes and, at finite coupling in a FM system, reproduce the structure of the phenomenological magnon–polaron theory used in Ref.~\cite{flebus17}. The key difference is that, here, both the existence of a well–defined chemical potential and the form of the response functions follow rigorously from rotationally invariant description of magnon–phonon coupling, rather than being imposed by analogy with purely magnonic transport. 

Although we have focused on centrosymmetric crystals in which transverse acoustic phonons carry no intrinsic angular momentum, our formalism extends naturally to systems in which lattice vibrations possess a well–defined angular momentum. In this case, the conserved $U(1)$ generator acquires an explicit lattice contribution, and the corresponding chemical potential necessarily couples to both spin and lattice degrees of freedom. Exploring this generalized structure would clarify how a lattice reservoir modifies the redistribution of angular momentum between spin and phonons, particularly in materials hosting chiral phonons or substantial symmetry breaking. 
Finally, the rotationally invariant framework developed here provides a consistent microscopic basis for thermodynamic and transport phenomena in regimes where magnon–phonon hybridization is strong and opens new opportunities to manipulate coupled spin and lattice angular–momentum currents via magnetic fields, symmetry engineering, and tailored phonon driving.

\section*{Supplemental Information}

Document S1. Supplemental Information, including additional derivations and discussion of dipolar interactions.

\section*{ACKNOWLEDGMENTS}
This work was supported by the National Science Foundation under Grant No. NSF DMR-2144086.

\bibliography{refs}
\end{document}


\pagestyle{empty}
\title{Supplemental Information \\[2pt]}

\maketitle

\section*{S1. Dipolar Interactions}
\setcounter{equation}{0}

Dipolar interactions provide a long-range, geometry-dependent contribution to spin–lattice coupling~\cite{Bloembergen1959},  microscopically distinct in origin from the short-range mechanisms considered in the main text. In this Supplemental Note, we derive the associated magnetoelastic terms and assess their relevance in the regime of interest, with particular emphasis on the ferromagnetic case, as in compensated collinear antiferromagnets the long-range dipolar field is strongly suppressed by sublattice compensation.

We start from the point-dipole Hamiltonian
\begin{equation}
\mathcal{H}_{\mathrm{dip}}=\frac{\mu_0 (\gamma \hbar)^2}{4\pi}\sum_{i<j} \left[\frac{\mathbf{S}_i\cdot \mathbf{S}_j}{|\mathbf{r}_{ij}|^3}-3\frac{(\mathbf{S}_i \cdot \mathbf{r}_{ij}) (\mathbf{S}_j \cdot \mathbf{r}_{ij})}{|\mathbf{r}_{ij}|^5}\right],
\label{eq:dipole1}
\end{equation}
where $\mathbf{r}_{ij}=\mathbf{r}_i-\mathbf{r}_j$, and the sum runs over all distinct pairs $i<j$. Decomposing the ionic positions as $\mathbf{r}_i=\mathbf{R}_i+\mathbf{u}_i$ yields the corresponding spin--lattice Hamiltonian:
\begin{equation}
\begin{split}
\mathcal{H}_\mathrm{dip}=\ 
&\frac{\mu_0 (\gamma \hbar)^2}{4\pi}\sum_{i<j} \left[\frac{\mathbf{S}_i\cdot \mathbf{S}_j}{|\mathbf{R}_{ij}+\mathbf{u}_{ij}|^3}\right.\\
&\left. -3\frac{[\mathbf{S}_i \cdot(\mathbf{R}_{ij}+\mathbf{u}_{ij})][\mathbf{S}_j\cdot(\mathbf{R}_{ij}+\mathbf{u}_{ij})]}{|\mathbf{R}_{ij}+\mathbf{u}_{ij}|^5}\right].
\end{split}
\label{eq:dipole2}
\end{equation}
To linear order in the relative displacement $\mathbf{u}_{ij}=\mathbf{u}_i-\mathbf{u}_j$, the Hamiltonian~\eqref{eq:dipole2} separates into a purely magnetic term:
\begin{equation}
\mathcal{H}_{\mathrm{dip}}=\frac{\mu_0 (\gamma \hbar)^2}{4\pi}\sum_{i<j}
\left[\frac{\mathbf{S}_i\cdot \mathbf{S}_j}{|\mathbf{R}_{ij}|^3}-3\frac{(\mathbf{S}_i \cdot \mathbf{R}_{ij}) (\mathbf{S}_j \cdot \mathbf{R}_{ij})}{|\mathbf{R}_{ij}|^5}\right],
\label{eq:dipoleMag}
\end{equation}
and a magnetoelastic contribution:
\begin{equation}
\begin{split}
\mathcal{H}_{\mathrm{mec}}
&=
\frac{\mu_0 (\gamma \hbar)^2}{4\pi}
\sum_{i<j} \left[
-3\frac{(\mathbf{R}_{ij}\cdot \mathbf{u}_{ij})(\mathbf{S}_i\cdot \mathbf{S}_j)}{|\mathbf{R}_{ij}|^5}\right.\\
&-3\frac{(\mathbf{S}_i\cdot \mathbf{R}_{ij})(\mathbf{S}_j\cdot \mathbf{u}_{ij}) +(\mathbf{S}_i\cdot \mathbf{u}_{ij})(\mathbf{S}_j\cdot \mathbf{R}_{ij})}{|\mathbf{R}_{ij}|^5}\\
&\left. +15\,\frac{(\mathbf{R}_{ij}\cdot \mathbf{u}_{ij})(\mathbf{S}_i\cdot \mathbf{R}_{ij})(\mathbf{S}_j\cdot \mathbf{R}_{ij})}{|\mathbf{R}_{ij}|^7}
\right].
\end{split}
\label{eq:dipoleMEC}
\end{equation}

It is easy to see that, although Eq.~\eqref{eq:dipole1} is anisotropic in spin space when the lattice is held fixed, the full spin--lattice Hamiltonian defined by Eqs.~\eqref{eq:dipoleMag} and \eqref{eq:dipoleMEC} is  rotationally invariant under a simultaneous global rotation of the spins and lattice vectors, since it is constructed entirely from scalar products of $\mathbf{S}_i$, $\mathbf{R}_{ij}$, and $\mathbf{u}_{ij}$. Similar to the anisotropy considered in the previous sections, the rotational invariance of the dipolar interaction is restored when consistently handling the spin and lattice degrees of freedom.
In the bulk limit relevant to the experiments motivating this work \cite{flebus17,kikkawa2016magnon,kikkawa22,Ramos19,Yang21}, dipolar interactions are governed by the standard magnetostatic kernel, which in the long-wavelength regime depends only on the propagation direction $\hat{\mathbf{k}}$ and therefore contributes a finite, angle-dependent term of order $\mathcal{O}(k^{0})$ to the magnon dispersion. This structure becomes explicit upon Fourier transforming the dipolar interaction in Eq.~\eqref{eq:dipole2} in the bulk limit, which  yields \cite{Damon1961,Harms2022}
\begin{align}
\mathcal{H}_{\mathrm{dip}}^{\mathrm{bulk}}
=
\frac{\mu_0 (\gamma \hbar)^2}{2 v_0}
\sum_{\mathbf{k} \neq 0}
S_{\mathbf{k}}^\alpha
\left(
\frac{k_\alpha k_\beta}{k^2}
-
\frac{\delta_{\alpha\beta}}{3}
\right)
S_{-\mathbf{k}}^\beta\;,
\end{align}
where $v_0$ is the unit cell volume. The magnetoelastic term Eq.~\eqref{eq:dipoleMEC} can be recast in an analogous bulk Fourier form:
\begin{align}
&\mathcal{H}_{\mathrm{mec}}^{\mathrm{bulk}}
=
\frac{i \mu_0 (\gamma \hbar)^2}{2 v_0 \sqrt{N}}
\sum_{\mathbf{k},\mathbf{q}}
u_{\mathbf{q}}^{\gamma}
S_{\mathbf{k}}^{\alpha}
S_{-\mathbf{k}-\mathbf{q}}^{\beta}
\nonumber \\
& \times
\left[
k_{\gamma}\frac{k_{\alpha}k_{\beta}}{k^2}
-
(k+q)_{\gamma}\frac{(k+q)_{\alpha}(k+q)_{\beta}}{|\mathbf{k}+\mathbf{q}|^2}
+
\frac{q_{\gamma}}{3}\,\delta_{\alpha\beta}
\right].
\label{Eq:FTdipoleMEC}
\end{align}
In the long-wavelength acoustic limit, the vertex in Eq.~\eqref{Eq:FTdipoleMEC} reduces to its leading-order expansion in $\mathbf{q}$, recasting the difference of dipolar kernels as a momentum-space derivative and yielding
\begin{align}
&\mathcal H_{\mathrm{mec}}^{\mathrm{bulk}}
=
-\frac{i\,\mu_0(\gamma\hbar)^2}{2v_0\sqrt N}
\sum_{\mathbf k,\mathbf q}
u_{\mathbf q}^{\gamma}\,
q_{\nu}\,
S_{\mathbf k}^{\alpha}
S_{-\mathbf k-\mathbf q}^{\beta}\nonumber \\
& \times \partial_{k_{\nu}}
\left[
k_{\gamma}
\left(
\frac{k_\alpha k_\beta}{k^2}-\frac{\delta_{\alpha\beta}}{3}
\right)
\right].
\label{969}
\end{align}
Equation~\eqref{969} shows explicitly that the magnetoelastic coupling induced by dipolar interactions is controlled by two distinct structures: the nonanalytic dipolar kernel, which depends only on the direction of $\mathbf{k}$, and the lattice deformation, which enters exclusively through the gradient of the displacement field. As a result, the leading magnetoelastic contribution scales as $q\,u$ (or equivalently $\nabla u$), matching the long-wavelength scaling of the local anisotropy-driven coupling, Eq.~(\textcolor{blue}{22}), considered in the main text. Thus, although microscopically distinct, the two mechanisms enter the effective low-energy theory on the same formal footing, so that the long-wavelength discussion of the main text depends only on their shared leading-order structure, rather than on the specific microscopic origin of the coupling.


\bibliography{refs}